# A Genetic Algorithm for the Calibration of a Micro-Simulation Model

Omar Baqueiro Espinosa

**Abstract**: This paper describes the process followed to calibrate a micro-simulation model for the Altmark region in Germany and a Derbyshire region in the UK. The calibration process is performed in three main steps: first, a subset of input and output variables to use for the calibration process is selected from the complete parameter space in the model; second, the calibration process is performed using a genetic algorithm calibration approach; finally, a comparison between the real data and the data obtained from the best fit model is done to verify the accuracy of the model.

## Introduction

An important step in the empirical study of a system using a discrete simulation approach (such as micro-simulation, agent-based or individual-based models) is the adaptation process. As part of adapting a model to a particular region, a set of input parameters representing the initial state of the represented system are determined. The majority of those parameters are usually defined using data derived from the analysis of the region; either from official statistics, from expert knowledge or sometimes making assumptions of the likely distributions that such parameters observe. In some cases, it is not possible to obtain a specific initial value for a subset of input parameters, either due to lack of data or because there is no prior knowledge about any correct assumption to define a value or a distribution for some input parameters. In these cases, it is possible to use an optimization process to derive a set of values for the input parameters which improve the *replicative validity* of the model; that is, aiming to minimize the difference between the data generated by the simulation and previously acquired data from the real system (Troitzsch, 2004).

This calibration process can be performed by several means. In principle, this procedure can be seen as an optimization problem: matching of independent variables (model inputs) to a set of dependent values (model outputs) for certain defined function (the model). Nevertheless, classical optimization tools such as regression may not be effective in finding a suitable combination of input parameters due to the inherent complexity of the variables interaction within the model. For this reason, other methods have been proposed which generate better input parameters for such type of models. Methods such as genetic algorithms (Li *et al.*, 1992) or Approximate Bayesian Computation (Lenormand *et al,*. 2011) have been previously used for model calibration with encouraging results.

In this work, we present a genetic algorithm used to calibrate a multi-parameter micro-simulation model developed during the European project PRIMA[1]. The calibration procedure here presented was used to calibrate an adapted version of this model for two study regions: The Altmark region in Germany and a region within Derbyshrie and Nottinghamshire in the United Kingdom. The presented work thus provides the results of an alternative calibration method to the one presented in (Lenormand, *et al.*, 2011).

---

[1] Prototypical policy Impacts on Multifunctional Activities in rural Municipalities. EU 7th Framework Research programme; 2008 – 2011; https://prima.cemagref.fr/the-project

The work is presented in 3 sections. The next section describes the selection of output indicators and the input parameters considered for the model calibration. Afterwards the structure of the used genetic algorithm is presented. Finally, the results from the optimization process for the two case study regions is presented and analyzed.

## Selection of Output and Input

The PRIMA model simulates the dynamics of virtual individuals living in a set of interconnected municipalities in a rural area. The dynamics of the model include demographic change (such as births, deaths aging, marriage, migration, and divorce), economic status change (student, worker, inactive, unemployed, retirement), and change of jobs (between a set of defined job types). The simulation evolves at yearly steps, with a starting year in the past (2000 for German region and 2001 for the UK region). A detailed description of the model is presented in (Huet and Deffuant, 2011).

To select the variables that will be used for the calibration of the model we can group the calibration in two parts. The first part concerns the calibration of the demographic outputs; the second part refers to employment/activity outputs.

### *Selection of output indicators for calibration*

To start, the output indicators to be used for calibration are selected. The selection is done by choosing the output variables for which real data is available for some years after the simulation starting point. The term *real data* here is used to comprise data acquired from national statistics which is assumed to represent the real state of the region being modelled. Due to limitations in the availability of real data from the regions, it is not possible to have all the necessary data for one specific year (for example, all data for 2006). Instead, the outputs of the model are compared to data available in different years (between 2001 and 2010) depending on the availability of real data. Nevertheless, as the simulation can provide data for every year simulated (from 2000 to 2020), the available real data can be compared with the simulation data of the corresponding year.

The selection of output indicators is then mainly driven by the availability of data. Table 1 presents the output indicators selected for comparison. The indicators are split in demographic and employment indicators to highlight the two aspects studied in the model.

**Table 1: Output indicators used for model calibration**

| Demography | | Employment | |
|---|---|---|---|
| Age structure | Municipality level. Yearly number of individuals grouped by ages. | Employment | Municipality level. Number of employed individuals grouped by age. |
| Births, deaths | Municipality level. Yearly number of births and deaths in the municipality | Unemployment | Municipality level. Number of individuals unemployment grouped by age. |
| Out-migration | Municipality level. Yearly number of people that move out of the municipality. | Sector of activity | District level. Proportion of individuals working in the different sectors (industries) |
| Household structure | District level. Percentage of households with 1, 2, 3, 4 or more persons. | Workplace | Municipality level. Number of individuals working in the municipality. |

**Source: Own table.**

For each of these indicators, real data for the modeled region is available. The available data describes some aspects of the region at different levels: although the majority of the available data is at municipality level, some data is only available at the district level. Because the model deals with measures at the municipality level, in case where corresponding real data is unavailable, percentage tables are used for comparison instead of absolute values.

For the cases were municipality data is available, the calibration is done against these indicators for each municipality in the model. An illustrative case is the household structure indicator, for which only district level data is available. In this case, the percentage of households with 1, 2, 3 and 4 or more persons at the district level (from the real data) is compared to the percentage of households for the municipalities of interest obtained by the simulation output.

## *Selection of input parameters*

Once the output indicators are chosen, it is time to select the input parameters that will be used for calibration. The values of these parameters will be modified (within certain limits) to optimize the fitness of the simulation results (measured by the output indicators previously selected) to the observed real data. The choice of the input parameters is done by first selecting input parameters for which values are unavailable in the real data used for initialization.

As it can be seen in Table 2, the values for some of the input parameters may be obtained from different data sources (such as the minimum age of having a child or the average number of children per couple). Such information is usually available as official statistics at aggregated levels (country or state level) and was used to define the value ranges for the optimization process.

**Table 2: Input Parameters (independent variables) used for model calibration.**

| Demography | | |
|---|---|---|
| **Paramater** | **Description** | **Range** |
| ageMinHavingChild | Minimum age required to have a child | [15, 20] |
| ageMaxHavingChild | Maximum age required to have a child | [40, 50] |
| nbChild | Average number of children per household | [1, 6] |
| probabilityTomakeCouple | Probabilty to accept joining a partner | [0, 0.05] |
| nbJoinTrials | Number of yearly trials done to look for a partner | [1, 50] |
| splittingProba | Yearly probability of splitting for a household | [0, 1] |
| probToAcceptNewResidence | Probability to accept a new residence when found | [0,1] |
| resSatisfactMargin | When changing residence, number of additional rooms available or needed tolerated. | [0,3] |
| **Employment** | | |
| **Paramater** | **Description** | **Range** |
| probStudyOutside | For students, probability of moving outside of the region to study higher education | [0,1] |
| probLookingRegionaljobs | Probability of looking for jobs outside the residence municipality | [0,1] |

**Source: Own table.**

To get a better understanding of the statistical behaviour of the model, correlation and linear regression and correlation analysis was performed. A total of 30000 simulation runs were executed initializing the input parameters with uniformly distributed random values. Results from running the simulations with the input values were used to compute the fitness of each set of inputs. The calculated correlation between each input parameter and Fitness value is shown in Table 3.

**Table 3: Correlation and $r^2$ between Input Parameters and Fitness Score**

| Input Parameter | Pearsons Corr. | $r^2$ |
|---|---|---|
| ageMinHavingChild | -0.071 | 0.005 |
| ageMaxHavingChild | 0.024 | 0.001 |
| nbChild | 0.038 | 0.001 |
| probabilityToMakeCouple | -0.090 | 0.008 |
| nbJoinTrials | -0.028 | 0.001 |
| splittingProba | 0.138 | 0.019 |
| probToAcceptNewResidence | 0.016 | 0.000 |
| resSatisfactMargin | 0.075 | 0.006 |
| probLookingRegionalJobs | -0.019 | 0.000 |
| jobVacancyRate | 0.865 | 0.748 |

**Source: Own table.**

Results from the analysis show that for the majority of the parameters, there is no strong correlation between them and the fitness score. As an exception, the parameter *jobVacancyRate* shows a strong positive correlation with the fitness score. This means that as the *jobVacancyRate* tends to one, the fitness of the model to the real data decreases (recall that a lower fitness value indicates simulation results replicating better the real data). Such behaviour is logical for the Altmark adaptation, given that the *jobVacancyRate* parameter defines the rate at which new jobs are created in the region at each time step (A *jobVacancyRate* of 1 would mean that the number of jobs is duplicated every year).

**Table 4: Linear regression analysis ($R^2$= 0.635)**

| | Coefficients | Standard Error | t Stat | P-value |
|---|---|---|---|---|
| Intercept | -7.56 | 2.06 | -3.66 | 0.0002 |
| ageMinHavingChild | 3.01 | 0.05 | 58.50 | 0.0000 |
| ageMaxHavingChild | -0.15 | 0.03 | -3.85 | 0.0001 |
| nbChild | -2.20 | 0.08 | -27.19 | 0.0000 |
| probabilityToMakeCouple | 13.61 | 0.39 | 34.64 | 0.0000 |
| nbJoinTrials | 0.05 | 0.008 | 6.39 | 0.0000 |
| splittingProba | -25.01 | 0.31 | -79.10 | 0.0000 |
| probToAcceptNewResidence | 7.81 | 0.23 | 33.18 | 0.0000 |
| resSatisfactMargin | -4.21 | 0.08 | -48.93 | 0.0000 |
| probLookingRegionalJobs | -5.65 | 0.38 | -14.64 | 0.0000 |
| jobVacancyRate | -53.73 | 0.2660 | -202.0192 | 0.0000 |

**Source: Own table.**

A meta-model was constructed using a linear-regression model. Results from the analysis are provided in Table 4; these show the importance of each variable in the outcome of the fitness value. Corroborating the correlation analysis, the linear model shows higher dependency shown by the *jobVacancyRate* input parameter.

# Model Calibration with Genetic Algorithm

The calibration procedure starts with the definition of the genetic algorithm used for the calibration process. The algorithm is characterized mainly by the structure of the chromosomes, the fitness function and the reproduction functions (including mutation, crossover and selection).

The chromosome is composed of the 11 parameters defined before in Table 2. At the beginning of the calibration process, a population of 50 chromosomes is initialized with random values uniformly assigned. Each parameter's initial value is limited to the range allowed for the same parameter.

Given a set of simulation outputs *O*, the fitness function is defined as the proportion of the distance between the real data and the obtained data from the simulation. For the fitness function, the difference between the simulated data and the real data is obtained with the function defined in Equation 1.

$$f = \sum_{i \in O} (\frac{x_i - x_r}{x_r})^2 \qquad \textbf{Equation 1}$$

Where $x_i$ is the output from the simulation and $x_r$ is the real data. The GA will minimize the value of *f*. This function considers the ratio of difference between the real and simulated data without being affected by the dimensionality of the data. Standardizing the fitness of each indicator is needed since the range of values of the output indicators varies.

The selection mechanism used is truncation selection (Crow and Kimura, 1979; Blickle and Thiele, 1996). The GA is run for 500 generations or until the fitness value reaches a long term plateau (more than 200 generations without a fitness increment). The data of the *fittest chromosome* (the combination of input parameters which provide the lowest value for the fitness function) is recorded. Evaluating the fitness of one chromosome requires running the simulation initializing the model with the chromosome's parameters. As the micro-simulation has stochastic components, the data resulting from the average of 5 simulation runs is considered for each set of input parameters. This average is considered as the output indicators used to calculate the fitness of a given chromosome.

The micro-simulation is left to run for 10 steps (from the year 2000 to 2010). This length of time allows obtaining simulated values comparable to the available real data. Each simulation run (with the 5 repetitions) required on average 40 seconds to execute (using an Intel Xeon CPU with 2.99 Ghz and 4 cores) by distributing each repetition in one CPU core. To reduce the time needed for the execution of the GA, a hash table was used to store previously scored chromosomes; this procedure was first proposed by (Povinelli and Feng, 1999).

## *Calibration Results*

As a result of the calibration procedure, a set of parameters providing a satisfactory model fit was obtained. The evolution of the fitness score obtained during the calibration of the Altmark region is shown in Figure 1. The optimization process reaches a local optimum of 16.76 in the fitness score from the generation 205.

**Figure 1: Model fitness throughout GA evolution for the Altmark adaptation**

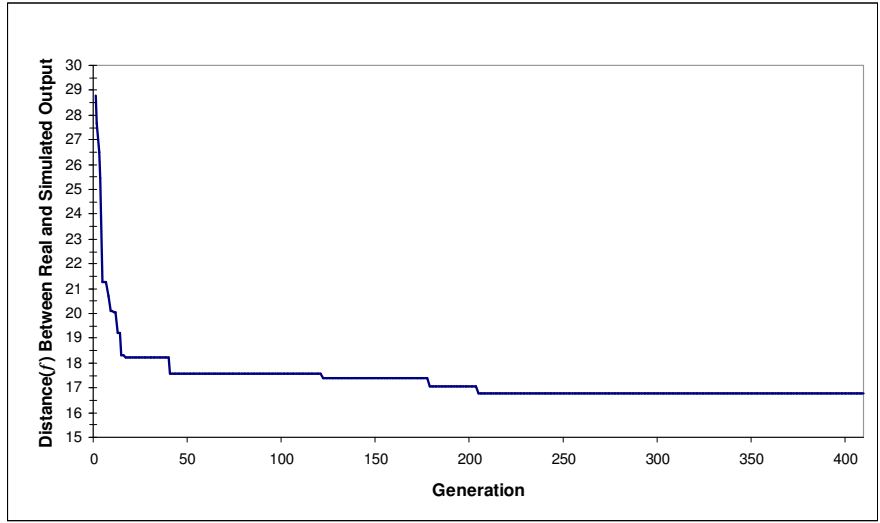

**Source: Own figure**

The resulting values for the input parameters (listed on Table 5) were tested by running further simulation experiments (with 100 repetitions). Results from simulations indicated a good mach for demographic indicators. Nevertheless, the trends of some economic indicators (such as unemployment, inactivity and retirement) were not accurately reproduced by the model (Baqueiro *et al.*, 2011). The inaccuracy was found to be caused by inadequate model assumptions on some of the dynamics of the adapted region.

**Table 5: Input parameters obtained after calibration**

| Parameter Name | Value after GA Optimization |
|---|---:|
| ageMinHavingChild | 19 |
| ageMaxHavingChild | 41 |
| nbChild | 2 |
| probabilityToMakeCouple | 0.0289 |
| nbJoinTrials | 19 |
| splittingProba | 0.124 |
| probToAcceptNewResidence | 0.0608 |
| resSatisfactMargin | 0 |
| probLookingRegionalJobs | 0.0575 |
| jobVacancyRate | 0.021 |

**Source: Own Table**

Examples of resulting simulation runs and their comparison with real data are provided in Figures 2, 3 and 4. It can be seen that for some cases, the trend produced by the calibrated model was close to the trend observed by the real data. In contrast, the unemployment trend could not be reproduced with the current model assumptions.

**Figure 2: Comparison of employment/ unemployment trend between Simulated and Real data.**

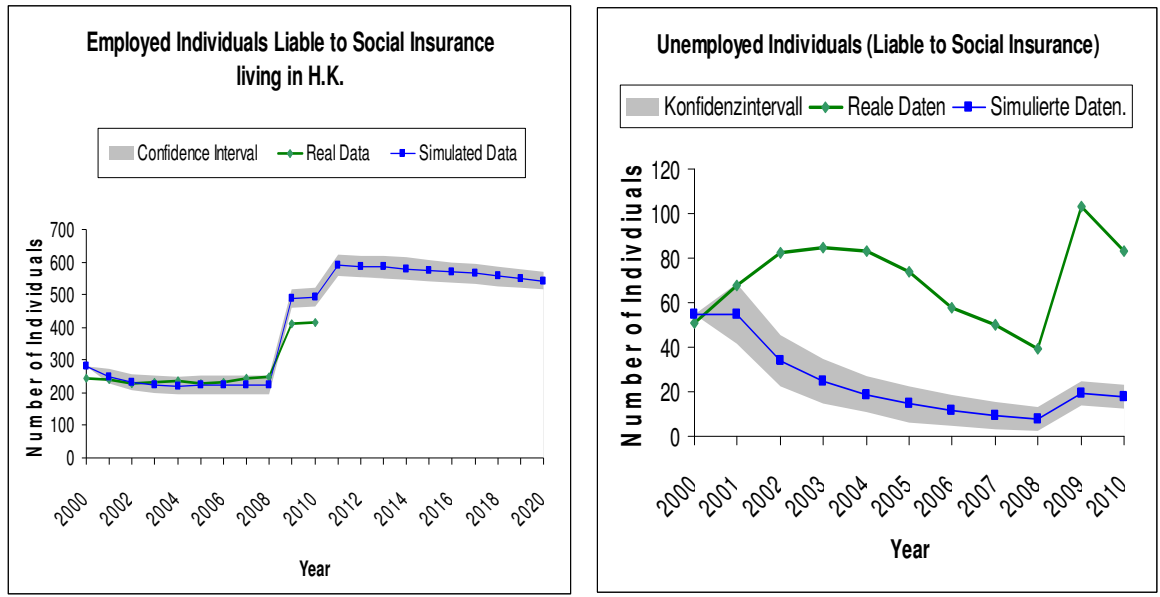

**Source: Own picture adapted from (Baqueiro *et al.,* 2011).**

**Figure 3: Comparison of birth/death rate evolution trend between Simulated and Collected data.**

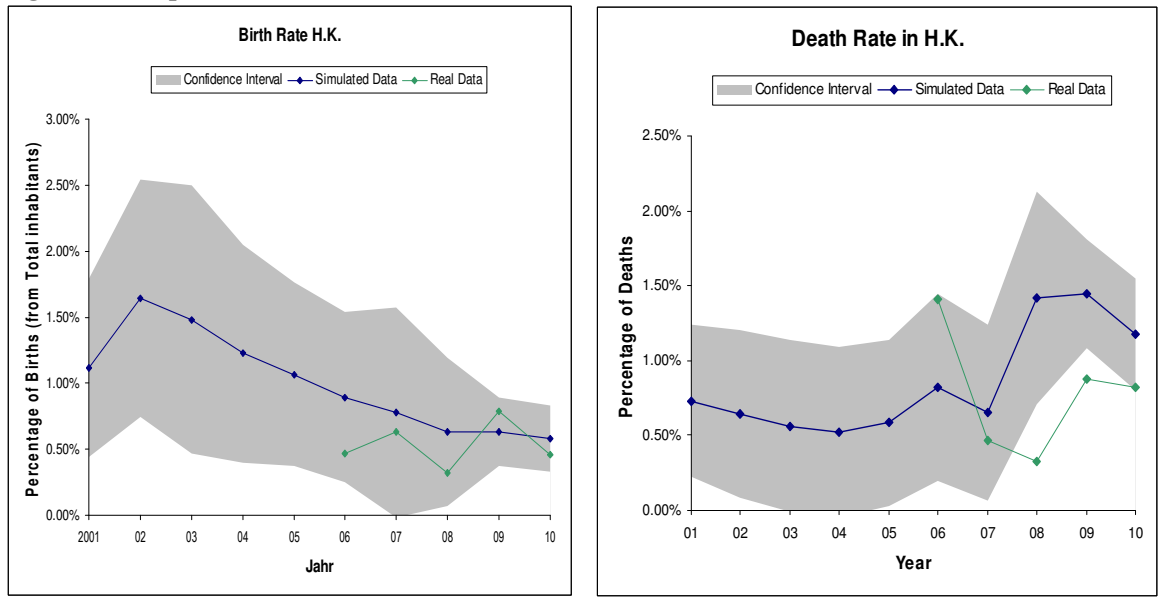

**Source: Own picture adapted from (Baqueiro *et al.,* 2011).**

**Figure 4: Comparison of sectors of activity evolution trend between Simulated and Collected data for the Hohenberg-Krusemark (Germany) region.**

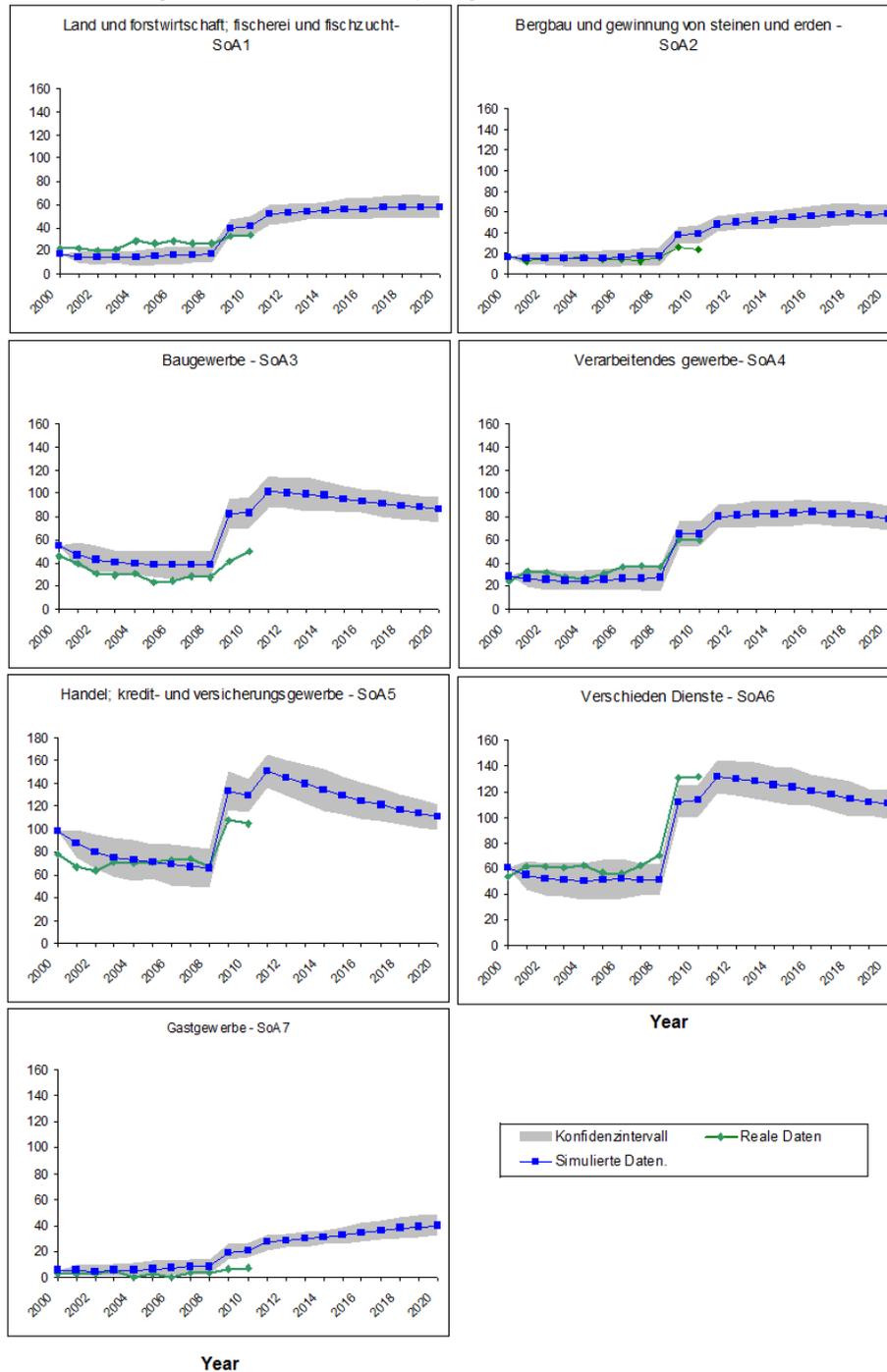

**Source: Own picture adapted from (Baqueiro *et al.*, 2011).**

# Conclusion

The present paper described the process followed to calibrate two regional adaptations of the PRIMA micro-simulation model. The calibration was performed for a selected subset of initialization parameters for which regional data was either unavailable at the required level or did not exist at all. Using a Genetic Algorithm, an adequate combination of the selected set of parameters is searched within a range of values which is reasonable for each parameter (the range itself obtained by empirical evidence obtained from literature or expert insight). Using the described process it was possible to obtain a set of input parameters which are both reasonable (within the logical ranges for each parameter) and which improve the model fidelity to the adapted region.

Two questions arise from this effort: First, the issue of whether it is possible to find a fitter chromosome providing better fidelity by modifying the GA parameters (initial population, selection and reproduction mechanisms). Although we believe that such an improvement may be possible, not much effort was concentrated in exploring such alternatives as it is presumed that the main issue preventing a better fitness is some assumptions in the model. Second, it may be possible to reduce the number of input parameters used to define the chromosomes. The correlation analysis showed there was high correlation between some pair of variables. Further analysis must be made to select a subset of input parameters from the current set.

An improved model calibration may be achieved after modifying the wrong underlying model assumptions. Additionally, the statistical analysis of the input and output space indicates that the optimization process may benefit from removing some of the input parameters (as the model output variation is explained by other parameters). Finally, this calibration approach considered mainly demographic variables while limiting the parameters related to economic status (*probLookingforRegionalJobs* and *jobVacancyRate*); this may be the cause of the performance of the model with regards to economic output indicators.